\documentclass[12pt,letterpaper]{article}
\usepackage[utf8]{inputenc}

\usepackage{graphicx,array}
\usepackage{url}
\usepackage{color}
\usepackage{latexsym}
\usepackage{amsthm}
\usepackage{amsmath}
\usepackage{amssymb}
\usepackage{amsfonts}
\usepackage[numbers,sort&compress]{natbib}
\usepackage{bm}
\usepackage{slashed}
\usepackage{bbm}
\usepackage{mathrsfs}
\usepackage{enumerate}
\usepackage{tikz}
\usepackage{siunitx}
\usepackage{mdframed}
\usepackage{setspace}  
\usepackage{esvect}
\usepackage{caption}
\usepackage{subcaption}

\usepackage{tcolorbox}%

\usepackage{hyperref} 
\hypersetup{
    colorlinks=true,       
    linkcolor=red,          
    citecolor=blue,        
    filecolor=magenta,      
    urlcolor=blue           
}
\usepackage[all]{hypcap} 
\usepackage{multirow}
\usepackage{multicol}
\usepackage{fancyhdr}
\usepackage{natbib}
\usepackage[version=3]{mhchem}

\newcommand{\nc}{\newcommand}

\nc{\beq}{\begin{equation}}
\nc{\eeq}{\end{equation}}
\nc{\beqa}{\begin{eqnarray}}  
\nc{\eeqa}{\end{eqnarray}}  
\nc{\bit}{\begin{itemize}}  
\nc{\eit}{\end{itemize}}  

\usepackage{floatrow}
\newfloatcommand{capbtabbox}{table}[][\FBwidth]

\usepackage{blindtext}

\title{ 
\vspace*{-2.3cm}
\begin{flushright}
\normalsize{N3AS-24-034}
\end{flushright}
\vspace{1.5cm}
 {\bf Radioactivity of Quark Nuggets}
\author{\large Yang Bai$^{\,a}$ and Mrunal Korwar$^{\,b,\,c}$}
\date{\small \it 
$^a$Department of Physics, University of Wisconsin-Madison, Madison, WI 53706, USA\\
$^b$ Department of Physics, University of California, Berkeley, CA 94720, USA \\
$^c$ Department of Physics and Astronomy, University of Kentucky, Lexington, KY 40506, USA
}
}

\begin{document}

\maketitle

\setlength{\parskip}{0.2ex}

\begin{abstract}
Quark nuggets $\ce{^A_ZQ}$, as Fermionic non-topological solitons, could have their mass per baryon smaller than ordinary nuclei and behave as exotic nuclei with different relations of atomic number and atomic mass number. Using both the degenerate Fermi gas model and the Friedberg-Lee shell model, we calculate the properties of quark nuggets made of up and down quarks. Similar to ordinary nuclei, quark nuggets could exhibit their own radioactivity, including gamma decay, beta decay, and (explosive) spontaneous fission, with the qualitative properties presented here. These quark nugget properties may provide guidance for searching for quark nuggets in situ from binary neutron star mergers.
\end{abstract}

\thispagestyle{empty}  
\newpage   
\setcounter{page}{1}  

\section{Introduction}
\label{sec:into}

A quark nugget is a baryonic state composed of quarks in the QCD de-confined phase. Hypothesized more than fifty years ago by A. R. Bodmer in Ref.~\cite{Bodmer:1971we}, it could have a mass per baryon smaller than that of ordinary nuclei. More generally, a quark nugget is a type of non-topological soliton with fermion constituents, extensively studied by T. D. Lee and collaborators in the 1970s~\cite{Lee:1974kn, Lee:1974uu, Friedberg:1976eg, Friedberg:1977xf}. These quark nuggets can also be considered “exotic elements” with atomic mass numbers far beyond the current periodic table. Discovering quark nuggets in our universe would not only identify the “ground state” of QCD but also help us understand the underlying constituents of dark matter, as emphasized by E. Witten in Ref.~\cite{Witten:1984rs}.

There is a long history of studies on quark nuggets in the literature. Treating quark nuggets as a degenerate Fermi state and introducing a “bag parameter” as the vacuum pressure to balance the quark Fermi pressure, Ref.~\cite{Farhi:1984qu} studied quark nuggets (with strange quarks) with both large and small baryon numbers. The stability of quark nuggets highly depends on the phenomenological bag parameter. Until now, we still do not have a robust prediction or measurement for this bag parameter, leaving the possibility of a stable quark nugget uncertain. More recently, Ref.~\cite{Holdom:2017gdc} used the linear sigma model plus constituent quarks (similar to the model by Friedberg and Lee in Refs.~\cite{Friedberg:1976eg, Friedberg:1977xf}) to model a quark nugget and found that quark nuggets could only contain up- and down-quarks, without strange quarks. Again, the stability of quark nuggets highly depends on the model’s parameters, and no robust statement can be made.

Quark nuggets may have been produced in the early universe through non-equilibrium processes such as first-order phase transitions~\cite{Witten:1984rs} or the collapse of domain walls~\cite{Liang:2016tqc}. Depending on the statistical environment after their formation, quark nuggets, as nontopological solitons, may undergo the solitosynthesis process, leading to a significant abundance in the current universe~\cite{Bai:2022kxq}. Since the QCD phase transition based on the Standard Model (SM) is a crossover~\cite{Aoki:2006we}, additional new physics beyond the SM is required to achieve a non-standard early-universe history around the QCD phase transition temperature.

Besides the possible early-universe production, astrophysical objects may also produce quark nuggets. A notable example is the neutron star with its potential quark-matter core. The merger events of two neutron stars or one neutron star with a black hole could produce quark nuggets~\cite{Witten:1984rs}. This can occur if the core of a neutron star already contains quark matter or if a phase transition happens during the merging process to form quark matter. A total mass of $\mathcal{O}(10^{-2}\,M_\odot)$ material~\cite{Pian:2017gtc} is anticipated to be ejected, including possible quark matter droplets or quark nuggets. One natural question is how we can confirm that quark nuggets are produced on-site from neutron star merger events.

The detection of ordinary heavy elements from neutron star mergers may teach us how to search for quark nuggets in an analogous way. As pointed out a while ago in Ref.~\cite{Eichler:1989ve}, the neutron star merger events can synthesize neutron-rich heavy elements by rapid neutron capture or the $r$-process. The observation of the binary neutron star merger GW170817~\cite{LIGOScientific:2017vwq} and its electromagnetic detection via GRB 170817A~\cite{LIGOScientific:2017zic} provides a confirmation of on-site heavy element productions from fitting a series of (atomic) spectra including redshift effects~\cite{Pian:2017gtc,Hotokezaka:2023aiq}.  

Motivated by the discovery of binary neutron star merger events and the understanding of the origin of heavy elements in our galaxy, we study the electromagnetic properties of quark nuggets. We anticipate that some radioactivity from quark nuggets could be detected through multi-messenger studies of future binary neutron star merger events. In this study, we will use a similar linear sigma model plus constituent quarks as in Ref.~\cite{Holdom:2017gdc} and choose a benchmark model point to ensure the two-flavor (up and down) quark matter contains stable states. For quark matter containing strange quarks, Ref.~\cite{Berger:1986ps} studied the radioactivity of strange quark matter based on the degenerate Fermi gas model with a bag parameter. Our study extends both Ref.~\cite{Holdom:2017gdc} and Ref.~\cite{Berger:1986ps} by adopting not only the degenerate Fermi gas model (see Ref.~\cite{Xie:2024mxr} for a recent study with different scalar potentials), but also the Friedberg-Lee model~\cite{Friedberg:1976eg, Friedberg:1977xf}, which provides a more precise modeling of quark nuggets, including various excited states. 

In our calculations of quark nugget properties, we consider both neutral and charged quark nuggets. We find that the minimum energy per baryon number, $E/A$, occurs at a non-zero charge, $Z_{\rm min}$, which scales as $A^{1/3}$. Around this $Z_{\rm min}$, there exists a range of stable quark nuggets, $\ce{^A_ZQ}$. Outside this stable range, quark nuggets undergo $\beta$ decay to reach stability. Besides $\beta$ decay, excited quark nuggets may have their quark constituents in excited states of the Friedberg-Lee shell model, requiring the solution of coupled Fermion and scalar equations of state. The transition to the ground state of quark nuggets results in the production of photons or gamma rays with energies ranging from MeV to 10 MeV. Additionally, we identify a spontaneous fission or “explosion” process in some lighter quark nuggets, whose baryon numbers lie between classically stable but quantum mechanically unstable limits. This process may lead to delayed explosions of quark nuggets, potentially manifesting as small afterglow fireworks following binary neutron star mergers.

Our paper is organized as follows. In Section~\ref{sec:basis}, we study the general properties of quark nuggets using two different models: the degenerate Fermi gas model in subsection~\ref{sec:deg-Fermi-gas-model} and Friedberg-Lee shell model in subsection~\ref{sec:FLSmodel}. We then list some main radioactivty of quark nuggets in Section~\ref{sec:radioactivity}: gamma decay~\ref{sec:gamma}, beta decay~\ref{sec:beta} and spontaneous fission~\ref{sec:fission}. We conclude our paper in Section~\ref{sec:discussion} and also briefly show the linear sigma model in Section~\ref{sec:LMSq}.  

\section{Properties of quark nuggets}\label{sec:basis}

In this section, we calculate some basic properties of quark nuggets based on two different models: the degenerate Fermi gas model and the Friedberg-Lee~\cite{Friedberg:1977xf} shell model. For both models, we use the linear sigma model with constituent quarks to interpolate different vacua. In addition to the linear sigma field $\Sigma$, the linear sigma model also contains the couplings of the sigma field to constituent quark fields $\mathcal{L}_{y} \supset -g (\bar{\psi}\Sigma \psi + h.c.)$ with $\psi=(\psi_u,\psi_d,\psi_s)$ as quark flavor fields. In this study, we fix the Yukawa coupling value $g=3.55$~\cite{Holdom:2017gdc}. More details about the linear sigma model can be found in Appendix~\ref{sec:LMSq}. 

We emphasize that the phenomenological linear sigma model serves only as a guide to study the qualitative properties of quark nuggets. The model parameters, especially the non-perturbative ones, are fixed by meson masses and decay constants. A robust determination of quark nugget properties requires a non-perturbative tool like Lattice QCD or a data-driven approach. The latter could be feasible if the core of a neutron star contains quark matter, which can be diagnosed based on the macroscopic properties of neutron stars. 

In the following, we choose two different models to study the quark nugget properties: Degenerate Fermi Gas Model and Friedberg-Lee Shell Model, with the latter providing a more fine-grained spectrum information as well as a more precise calculation for a quark nugget with a smaller baryon number. 

\subsection{Degenerate Fermi gas model}
\label{sec:deg-Fermi-gas-model}
In the degenerate Fermi gas model, one treats quarks inside quark nuggets as a degenerate fermion gas without tracking the individual quark wave functions. This model focuses on the mean or average behaviors of quarks or using the Thomas-Fermi approximation. In the context of ``abnormal nuclear states," discussed a while ago by T. D. Lee and collaborators~\cite{Lee:1974ma,Lee:1974kn,Lee:1974uu}, the finite-density effects of fermions can modify the expectation value of the sigma field and lead to approximately massless fermions inside the abnormal states. The energy per baryon number, or the stability of quark nuggets, relies on the competition between the increase in the vacuum energy of the sigma field and the decrease in the quark Fermi energy.

Ignoring the small isospin-violating vacuum expectation value (VEV), one can choose a basis to parametrize the diagonal fields of the sigma field as $\Sigma = \mbox{diag}\{\sigma_n/2, \sigma_n/2, \sigma_s/\sqrt{2} \}$. Both the VEVs of $\sigma_n$ and $\sigma_s$ can be influenced by the finite density effects of fermions. As emphasized in Ref.~\cite{Holdom:2017gdc}, quark nuggets may not contain strange quarks due to the finite mass of strange quarks and the relatively low Fermi momentum in the ground state of quark nuggets. In our study, we choose a benchmark model point in \eqref{eq:benchmark-model} from Ref.~\cite{Holdom:2017gdc} to demonstrate some characteristic properties of the two-flavor (up- and down-) quark nuggets. For the calculation of radioactivity of strange quark nuggets based on the degenerate Fermi gas model, we refer the reader to Ref.~\cite{Berger:1986ps} for an earlier study. 

The total energy density $\mathcal{E}$ of a quark nugget contains several parts: the vacuum energy density and field-gradient energy density $\mathcal{E}_{\sigma}$ of the sigma fields; the Fermi energy density of quarks (and electron) $\mathcal{E}_{\rm F}$; the Coulomb energy density $\mathcal{E}_{\rm C}$. In this analysis, we ignore the energy contained in gluons, which is subleading and can be absorbed into some phenomenological parameters in the sigma field effective potential~\cite{Berger:1986ps}. For a spherically-symmetric quark nugget, the summed energy density is 
\beqa
\mathcal{E}(r) = \mathcal{E}_{\sigma}(r) + \mathcal{E}_{\rm F}(r) + \mathcal{E}_{\rm C}(r) ~,
\eeqa
with 
\beqa
\mathcal{E}_{\sigma}(r) &=& V\big[\sigma_{n}(r),\sigma_{s}(r)\big] - V\left(v_{n},v_{s}\right) +  \frac{1}{2}\big[\nabla \sigma_n(r)\big]^{2} +  \frac{1}{2}\big[\nabla \sigma_s(r)\big]^{2} ~, \\
\mathcal{E}_{\rm F}(r) &=& \sum_{i = u, d, s, e} \frac{2\, c_i}{(2\pi)^3}\, \int^{p_{F, i}(r)}_0\,d^3p \,\sqrt{p^2 + m_i^2\big[\sigma_n(r), \sigma_s(r)\big]} ~, \\
\mathcal{E}_{\rm C}(r) &=& n_Z(r)\,\frac{\alpha}{r}\,\int^r_0 dr'\,4\pi\,r'^{2}\,n_Z(r')~.
\eeqa
Here, $c_i =3$ for quarks and $c_i = 1$ for electron. $v_n$ and $v_s$ are VEVs of $\sigma_n$ and $\sigma_s$ in the normal vacuum, respectively. $v_n = 92$~MeV and $v_s = 90.5$~MeV for the benchmark model point. The field-dependent masses of constituent quarks, $m_u = m_{u, 0} + g\,\sigma_n(r)$, $m_d = m_{d, 0} + g\,\sigma_n(r)$ and $m_s = m_{s, 0} + \sqrt{2}\,g\,\sigma_s(r)$ with $g = 3.55$ and the bare quark masses $m_{u, 0} = m_{d, 0} = 3.5$~MeV and $m_{s, 0} = 92$~MeV. $p_{F, i}(r)$ represents the Fermi momentum for different fermions with the Fermion number density $n_i(r) = c_i\,p^3_{F, i}(r)/(3\pi^2)$. The electric charge density $n_Z(r) = \sum_i q_i \,n_i(r)$ with the Fermion electric charges $q_u = +2/3$, $q_d = q_s = -1/3$ and $q_e = -1$. $\alpha = 1/137$ is the fine-structure constant. The baryon number density is $n_A(r) = \frac{1}{3}[n_u(r) + n_d(r) + n_s(r)]$ with the total baryon number $A = 4\pi\,\int dr r^2 n_A(r)$. The total electric charge is $Z = 4\pi\,\int dr r^2 n_Z(r)$. The total energy or mass of a quark nugget is $E = 4\pi\,\int dr r^2 \mathcal{E}(r)$. 

\subsubsection{The bulk Limit}
\label{sec:bulk-limit}

For a large baryon number $A$, the size of the nugget is much larger than the surface, and hence the surface effects can be neglected. Consequently, $\sigma_{n, s}(r)$ and $p_{F}(r)$ can be considered constant for $r<R$, where $R$ is the radius of a quark nugget. Defining $f_i \equiv (p_{F,i}/p_F)^3$ with $p_{F}=(3\pi^{2} n_{A})^{1/3}$ as the Fermi momentum associated with the baryon number density $n_{A}$, one has $f_i = (n_i/3)/n_A$ as the fraction of an individual quark’s contribution to the baryon number density, and $f_u + f_d + f_s  = 1$.

For a quark nugget without electrons, the charge neutralization can be satisfied by requiring $n_Z(r) = 0$ for $r < R$ or $\frac{2}{3} f_u - \frac{1}{3} f_d - \frac{1}{3} f_s = 0$. We first consider a two-flavor quark nugget with no strange quark ($f_s = 0$) and will come back to the possibility of strange quark nuggets later. For this case, electrically-neutral quark nuggets satisfy $f_u = \frac{1}{3}$ and $f_d = \frac{2}{3}$. The $\sigma_s(r)$ field stays the same inside and outside the quark nugget, or $\sigma_s(r) = v_s$. The $\sigma_n(r)$ field, on the other hand, can have a different value $\sigma_n$ (taken to be a constant) inside from the outside value $v_n$. The energy density contains both the potential energy and the Fermi energy and is just a function of $\sigma_n$ and $p_F$, or $\mathcal{E}(\sigma_n, p_F)$. 

\begin{figure}[ht!]
    \centering
        \includegraphics[width=0.47\linewidth]{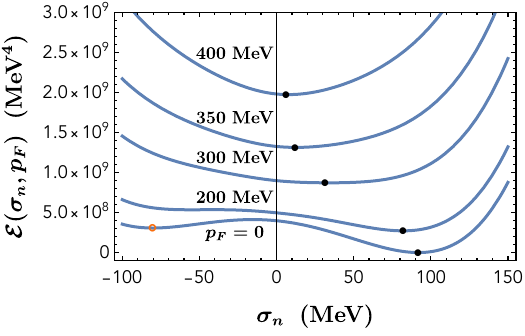} \hspace{3mm}
        \includegraphics[width=0.45\linewidth]{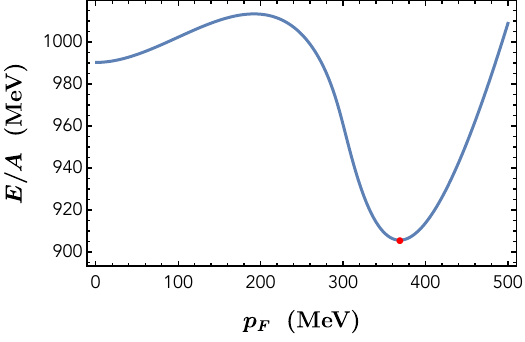} 
    \caption{\textit{Left panel}: The energy density as a function of the sigma field value $\sigma_n$ for different baryon Fermi momenta $p_F$ in the bulk limit. The black dots denote the minimum locations of the energy density. \textit{Right Panel}: The energy per baryon as a function of the baryon Fermi momentum $p_F$ after minimizing the energy density in $\sigma_n$. The minimum (red dot) is located at $p_F = 368.7 \, \mathrm{MeV}$ with $E/A|_{\rm min} = 905.6 \, \mathrm{MeV}$.
}
    \label{fig:bulk-energy}
\end{figure}

In the left panel of Fig.~\ref{fig:bulk-energy}, we show the energy density as a function of $\sigma_n$ for different values of $p_F$. One can see that the minimum location of $\sigma_n$ decreases as $p_F$ increases. For a large value of $p_F$ around 400 MeV, the value of $\sigma_n$ becomes close to zero and demonstrates the chiral symmetry restoration inside a dense quark nugget. In the right panel of Fig.~\ref{fig:bulk-energy}, we show the energy per baryon, or $E/A = \mathcal{E}/n_A$, as a function of $p_F$ after setting $\sigma_n$ to be the value that minimizes $\mathcal{E}$. As a function of $p_F$, $E/A$ shows a non-monotonic behavior with the minimum value located at $\sigma_n = 8.55$ MeV and $p_F = 368.7$ MeV with $E/A|_{\rm min} = 905.6$ MeV, which is smaller than the energy per baryon of the iron element $M_{\rm Fe}/56 \approx 929$ MeV.

A similar calculation can be performed by including the strange quark component. Because of a heavier bare strange quark mass, the baryon number Fermi momentum is required to be above 600 MeV to restore the chiral symmetry or to decrease the field value of $\sigma_s$ inside the quark nugget. As a result, the energy per baryon is above 1000~MeV for the current benchmark point. Therefore, the strange quark nugget may not be energetically preferred compared to the two-flavor quark nugget (see Ref.~\cite{Holdom:2017gdc} for emphasizing this point). Although this point is model-dependent and relies on the model parameters in the phenomenological linear sigma model, we take this point of view and choose the current model benchmark point to study the radioactivity of two-flavor quark nuggets.

It is also useful to compare the $E/A|_{\rm min}$ with the result using a simpler approach in Ref.~\cite{Witten:1984rs}, where a ``bag parameter" $B$ is introduced to model the vacuum pressure. For the linear sigma model, one could identify the scalar potential energy at $\sigma_n = 0$ as the bag parameter or 
\beqa
B \equiv V(\sigma_n = 0, v_s) - V(v_n, v_s) ~. 
\eeqa
For the current benchmark, we have $B = (141.2\,\mbox{MeV})^4$. Following the simple calculation in Ref.~\cite{Witten:1984rs}, we have the quark nugget energy per baryon as 
\beqa
[E/A]^{\rm bag} = 3\sqrt{2\pi}\,(f_u^{4/3} + f_d^{4/3})^{3/4}\,B^{1/4} ~,
\eeqa
which is 909.3~MeV for $f_u=1/3$ and $f_d = 2/3$ and close to the value obtained previously. The radius of the quark nugget has $R = 3^{1/3}\,(1 + 2^{4/3})^{1/4}\,B^{-1/4}\,A^{1/3}/[2(2\pi)^{1/6}] \approx 0.72\times B^{-1/4}\,A^{1/3}$. 

To neutralize the electric charge, one can also include electrons inside the quark nuggets. Obviously, this adds an additional electron Fermi energy into the system without increasing the baryon number. So, this is anticipated to be energetically not preferable. Defining $p_{F,e} = f_e\,p_F$, the charge neutrality requires $2 f_u - f_d - f_e = 0$. Repeating the same calculation as the case without electrons, we find $E/A|_{\rm min} = 905.5 \, \mathrm{MeV}$ with $\sigma_n = 8.54 \, \mathrm{MeV}$, $p_F = 368.7 \, \mathrm{MeV}$, and $f_e = 0.0055$ ($f_u = \frac{1}{3} + \frac{f_e}{3}$ and $f_d = \frac{2}{3} - \frac{f_e}{3}$). This means that including electrons to neutralize the electric charge has a negligible effect on the energy per baryon of quark nuggets.

For a charged quark nugget without electrons, one can relax the condition of $f_u = 2 f_d$ with the addition of the Coulomb energy. For a uniform charge density within a radius $R$, the Coulomb energy is 
\beqa
E_{\rm C} = \frac{3\,\alpha\,Z^2}{5\,R} = \frac{3\,\alpha\,A^2}{5\,R}\, (f_d - 2\,f_u)^2 ~.
\eeqa
Including the Fermi and potential energies, the total energy of a quark nugget is 
\beqa
E = \frac{9 \left(\frac{3}{2}\right)^{2/3} \pi^{1/3}\,A^{4/3}
   \left(f_d^{4/3}\,+\,f_u^{4/3}\right)}{4\,R}\,+\,\frac{3\,
   \alpha\,A^2\,(f_d \,-\,2 f_u)^2}{5\,R}+\frac{4}{3} \pi\,R^3\, B ~.
\eeqa
Minimizing the energy in $f_d = 1 - f_u$ and $R$, the minimum of $E$ prefers a nonzero electric charge as [in the limit of $A\gg \mathcal{O}(\alpha^{-2/3}) \sim \mathcal{O}(1000)$]
\beqa
\label{eq:bulk-Z-min}
Z_{\rm min} \, =\, \frac{5(2^{1/3} - 1)\,\pi^{1/3}\,A^{1/3}}{2\times 6^{2/3}\,\alpha}~.
\eeqa
The energy of a quark nugget $\ce{^A_Z Q}$ for $Z=Z_{\rm min}$ is 
\beqa
E(A, Z_{\rm min}) = (1 + 2^{4/3})^{3/4}\,\sqrt{2\pi}\,A\,B^{1/4} - \frac{5(2^{4/3} + 2^{2/3} - 4)\pi^{5/6}}{2^{17/6}\,3^{2/3}\,(1 + 2^{4/3})^{1/4}}\,\frac{A^{1/3}}{\alpha}\,B^{1/4} ~.
\eeqa

Around $Z_{\rm min}$, the quark nugget energy has the following quadratic dependence in $Z - Z_{\rm min}$
\beqa
E(A, Z) = E(A, Z_{\rm min}) + \frac{1}{2}\, \frac{4\times 3^{2/3}\,(2\pi)^{1/6}\,\alpha}{5(1+2^{4/3})^{1/4}\,A^{1/3}}\,B^{1/4}\,(Z - Z_{\rm min})^2~,
\eeqa
which means that in the large $A$ limit a wide range of quark nuggets with $Z$ around $Z_{\rm min}$ are stable against $\beta^\pm$ decays. For the range of charges without $\beta^\pm$ decay or $Z > Z_{\beta^-}$, quark nuggets can have the electron capture process to reduce its positive electric charge: $\ce{^A_Z Q} + e^- \rightarrow \ce{^A_{Z-1} Q} + \nu_e$ by emitting a low-energy neutrino. 

\subsubsection{The surface properties}
\label{sec:surface}
For a smaller quark nugget, its surface energy becomes important and can not be ignored. The scalar field profile $\sigma_n(r)$ and the quark Fermi momenta $p_{F, u}(r)$ and $p_{F, d}(r)$ become radius-dependent. For instance, the equation of motion of $\sigma(r)$ becomes
\beqa
\label{eq:sigma-EOM-surface}
\nabla^{2}\sigma_{n}(r) &=& \frac{\partial V[\sigma_n(r)]}{\partial \sigma_{n}} + \frac{\partial \mathcal{E}_F}{\partial \sigma_n}[\sigma_n(r), p_{F, i}(r)] ~,
\eeqa
which requires the determination of $p_{F, i}(r)$ from minimizing $E/A$. Instead of obtaining the exact function for $p_{F, i}(r)$, one can use the variation method to obtain the groundstate of a quark nugget with a fixed value of $A$. More specifically, we make the following ansatz profile for the Fermi momentum 
\beqa
\label{eq:pfr}
    p_{F}(r) = p_{F}^{\rm{bulk}}\,\times\, \frac{1}{2} \Big{[}1 - \tanh\big{(}\frac{r-R}{a}\big{)} \Big{]} ~,
\eeqa
which introduces one parameter $R$ to denote the radius and one parameter $a$ to denote the surface thickness. We fix the overall coefficient to be the bulk-region Fermi momentum $p_F^{\rm bulk}$, which is $368.7$~MeV for the benchmark model point. As in the previous bulk limit, we assume that both quark Fermi momenta have the same spatial profiles such that $p_{F, u}(r) = f_u\,p_F(r)$ and   $p_{F, d}(r) = f_d\,p_F(r)$ with $f_u + f_d = 1$. We do not include electrons in the analysis, but will analyze both neutral and charged quark nuggets. 

In the ``thin-wall" limit with $a \ll R$, the total baryon number has the approximate formula
\beqa
    A \approx \frac{4}{9\pi} (p_{F}^{\rm{bulk}}\,R)^{3}\, \Big{[]}1 - \frac{9\,a}{4\,R} + \frac{(3+\pi^{2})\,a^{2}}{4 \,R^{2}} - \frac{3\pi^{2}\,a^{3}}{16\,R^{3}}\Big{]} ~.
\eeqa
For a value of $R$ and $a$, we solve the equation of motion for $\sigma_n$ in Eq.~\eqref{eq:sigma-EOM-surface} and then calculate $E/A$. We have found that the minimum of $E/A$ prefers a constant value of $a$ for different choices of $R$. In other words, different quark nuggets with different $A$ have similar surface thicknesses. 

\begin{figure}[ht!b]
    \centering
        \includegraphics[width=0.49\linewidth]{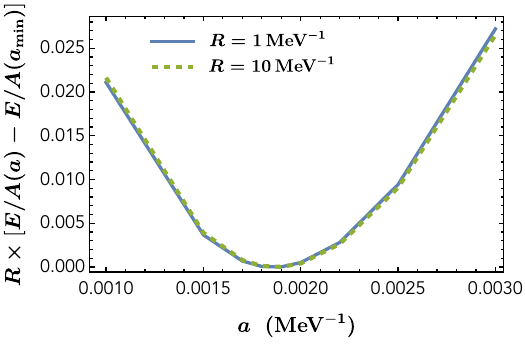} \hspace{3mm}
        \includegraphics[width=0.44\linewidth]{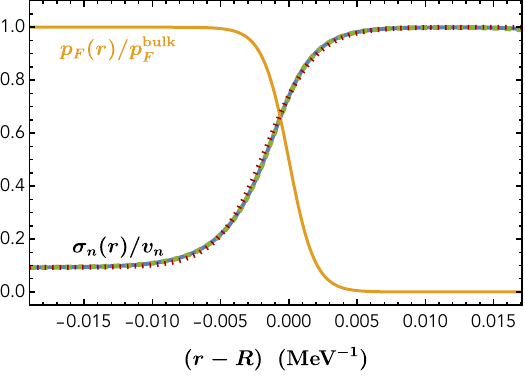} 
    \caption{{\it Left panel}: The distribution of $E/A$ in terms of the surface thickness parameter $a$ for two different choices of $R$. Both cases have the minimum located at $a_{\rm{min}}=0.0019\,\mbox{MeV}^{-1}$ with $E/A(a_{\rm min}) = 906.10 \, \rm{MeV}$ for $R=1 \, \rm{MeV}^{-1}$ and $905.65 \, \rm{MeV}$ for $R=10 \, \rm{MeV}^{-1}$. {\it Right panel}: The normalized profiles of $\sigma_{n}(r)/v_n$ and $p_{F}(r)/p_F^{\rm bulk}$ around the boundary region for $a = a_{\rm{min}}$. The profiles for $R=1 \, \rm{MeV}^{-1}$ (blue solid) and $10\, \rm{MeV}^{-1}$ (green dashed), as well as the fitted function (red dotted) from \eqref{eq:sigma_n-fit-function}, are overlapping with each other.}
    \label{fig:surfaceprofile}
\end{figure} 

For neutral quark nuggets with $f_u = 1/3$ and $f_d = 2/3$, and for $R = 1\,\mbox{MeV}^{-1}$ and $10\,\mbox{MeV}^{-1}$, we show the distributions of $E/A$ as a function of $a$ around the minimum value of $E/A$ in the left panel of Fig.~\ref{fig:surfaceprofile}. Both distributions have the minimum located at $a_{\rm min} = 0.0019\,\mbox{MeV}^{-1}$. This confirms the profile similarity of different quark nuggets. In the right panel of Fig.~\ref{fig:surfaceprofile}, we show the normalized profiles of $\sigma_n(r)/v_n$ and $p_F(r)/p_F^{\rm bulk}$ for $a = a_{\rm min}$ around the boundary or $r$ around $R$. Both cases with $R = 1\,\mbox{MeV}^{-1}$ and $10\,\mbox{MeV}^{-1}$ have identical profiles around the boundary. We have also found that the following formula provides a good fit to the $\sigma_{n}(r)$ profile.
\beqa
\label{eq:sigma_n-fit-function}
    \sigma_{n}(r) = \sigma_{n}^{\rm{bulk}} + (v_{n}-\sigma_{n}^{\rm{bulk}}) \frac{1}{2} \Big{[}1 + \tanh\left(\frac{r-R+0.00165}{0.0036}\right)\Big{]} \,.
\eeqa
Here, $v_{n}=92\,\rm{MeV}$; $\sigma_{n}^{\rm{bulk}}=8.547 \, \rm{MeV}$; both $r$ and $R$ are in $\mbox{MeV}^{-1}$.  

For charged quark nuggets, we do not fix $f_d = 1- f_u$ but include the Coulomb energy in the system. Repeating the similar calculation as the neutral one, we show $E/A$ as a function of $A$ in the left panel of Fig.~\ref{fig:EbyA_compare}. In the right panel of Fig.~\ref{fig:EbyA_compare}, we also show the value of $Z_{\rm min}$ as a function of $A$. In the large $A$ limit, the dependence of $Z_{\rm min}$ with respect to $A$ agrees very well with the analytic formula in \eqref{eq:bulk-Z-min} based on the bulk limit calculation.

\subsection{Friedberg-Lee shell model}
\label{sec:FLSmodel}

For small $A$, the Friedberg-Lee model~\cite{Friedberg:1976eg, Friedberg:1977xf} provides a better or more precise estimation for the quark nugget energy, as well as the study of the radioactivity of quark nuggets. In this model, one solves the coupled Fermion and scalar equations of motion to obtain various Fermion energy levels. For a given baryon number $A$, one could fill the allowed energy levels subject to the Pauli exclusion principle, taking into account the quark color, spin, and flavor factors. In this section, we first adopt an approximation by assuming a step-function profile for the scalar field (motivated by the degenerate Fermi gas model) to analytically obtain the Fermion energy levels.

\subsubsection{Step-function approximation for the scalar profile}
\label{sec:FLSmodel}

In this approximation, we will assume the scalar field or the $\sigma_n$ field has a step-function profile. The $\sigma_n$ field has a zero field value inside a quark nugget radius $R$ and the ordinary VEV $v_n$ outside $R$. With this assumption, the Fermion or quark field has different masses inside and outside. The Fermion energy levels can be analytically solved and can be used to study the qualitative features of radioactivity of quark nuggets. 

We start with the Dirac equation for the quarks 
\begin{equation}
    \left[i \gamma^{\mu}\partial_{\mu} - m_{q}(r)\right]\psi(t,\bm{r}) = 0\, .
\end{equation}
Here, in the Dirac's representation of the gamma matrices, one has $\gamma^{0}=\beta=\text{diag}(\mathbbm{I}_{2}, -\mathbbm{I}_{2})$ and $\gamma^{i}=\beta \alpha^{i}$ with $\alpha^{i}=\{\{0,\sigma^{i}\},\{\sigma^{i},0\}\}$ with $\sigma^i$ denoting the Pauli matrix. For a spherically symmetric effective mass, the general solution is 
\beqa
\label{eq:fermion-decomposition}
\psi_{nljm}(\bm{r}) =
\begin{pmatrix}
i\,f_{nlj}(r)\,\mathcal{Y}_{ljm}(\hat{\bm{r}} ) \\
g_{nlj}(r)\,\vec{\sigma}\cdot \hat{\bm{r}} \mathcal{Y}_{ljm}(\hat{\bm{r}} )
\end{pmatrix}
=
\begin{pmatrix}
i\, \frac{a_{nlj}(r)}{r} \,\mathcal{Y}_{ljm}(\hat{\bm{r}} ) \\
\frac{b_{nlj}(r)}{r}\,\vec{\sigma}\cdot \hat{\bm{r}} \mathcal{Y}_{ljm}(\hat{\bm{r}} ) 
\end{pmatrix}
~,
\eeqa
with $\psi_{nljm}(t,\bm{r}) = \psi_{nljm}(\bm{r}) e^{-i E t}$.
Here, $\mathcal{Y}_{ljm}(\hat{\bm{r}})$ is the two-component Pauli spinor harmonic with
\beqa
\mathcal{Y}_{ljm}(\hat{\bm{r}}) = \sum_{m_l,\,m_s} \langle lm_l; \frac{1}{2} m_s | j m \rangle\,Y_{l m_l}(\hat{\bm{r}}) \,\chi_{m_s} ~. 
\eeqa
Here, $\langle \cdots \rangle$ represents Clebsch–Gordan (CG) coefficients. There is a useful relation: $\vec{\sigma}\cdot \hat{\bm{r}} \mathcal{Y}_{ljm}(\hat{\bm{r}} )  =  \mathcal{Y}_{\ell jm}(\hat{\bm{r}} )$ with $\ell = l \mp 1$ when $l = j \pm 1/2$, as well as $(\vec{\sigma}\cdot \hat{\bm{r}})^2 = \mathbb{I}_2$.

The coupled equations for $a_{nlj}(r)$ and $b_{nlj}(r)$ are 
\beqa
\label{eq:ar-br-coupled}
\frac{da_{nlj}(r)}{dr} &=& \frac{\kappa_{lj}}{r} a_{nlj}(r) + [m_{q}(r) + E]\,b_{nlj}(r) ~, \nonumber\\
\frac{db_{nlj}(r)}{dr} &=&   [m_{q}(r) - E]\,a_{nlj}(r) - \frac{\kappa_{lj}}{r} b_{nlj}(r) ~.
\eeqa
Here, $\kappa_{lj} = \frac{1}{2} \pm ( l + \frac{1}{2})$ for $j = l \pm \frac{1}{2}$. Combining the two first-order differential equations, and in the special case of constant mass, one has the second-order differential equation for $a_{nlj}(r)$ as 
\beqa
\label{eq:a(r)-constant-mass}
a''_{nlj}(r) - \frac{l (l + 1)}{r^2}\,a_{nlj}(r) + [E^2 - m_{q}^2]\,a_{nlj}(r) = 0 ~.
\eeqa
The general solution to the above equation is 
\beqa
a_{nlj}(r) = c_1 \, r\,j_l\left(\sqrt{E^2 - m_{q}^2} \, r \right) + c_2 \, r\,y_l\left(\sqrt{E^2 - m_{q}^2} \, r \right) ~,
\eeqa
with $j_l(x)$ and $y_l(x)$ as the spherical Bessel functions. 
For a step-function mass for quarks
\beqa
m_{q}(r) =  \left\{   \begin{array}{l l}  
0 ~, & \quad r \leq R ~, \\ [0.0em]
m_0 \equiv g\,v_n ~, & \quad r > R ~.
  \end{array}
\right.
\eeqa
Here, we have neglected the small contributions from $m_{u, 0}$ and $m_{d, 0}$ and considered $\sigma_{n}(r) \rightarrow 0$ for $r < R$ or a complete chiral symmetry restoration inside the quark nugget. In this case, 
the solution to the Dirac equation is 
\beqa
a_{nl}(r) = \left\{   \begin{array}{l l} 
a^{\rm in}_{nl}(r) = d_1 \,r\,j_l(E_{n l} \, r) ~, & \quad r \leq R ~, \\ [0.5em]
a^{\rm out}_{nl}(r) = d_2 \,r\, [j_l(i\,\sqrt{m_0^2 - E^2_{n l} } \, r) + i\,y_l (i\,\sqrt{m_0^2 - E^2_{n l} }\, r) ] ~, &\quad r > R ~.
  \end{array}
\right.
\eeqa
Here $E_{nl}$ is the bound state energy. Imposing the boundary conditions to have a continuous wave function and its derivative at  $r=R$ leads to the quantization condition \begin{equation}\label{eq:energylevels}
    \frac{E_{nl}\,j_{l+1}(E_{nl}\,R)}{j_{l}(E_{nl}\,R)} = \sqrt{m_{0}^{2}-E_{nl}^{2}}\,\frac{K_{l+3/2}(\sqrt{m_{0}^{2}-E_{nl}^{2}}\,R)}{K_{l+1/2}(\sqrt{m_{0}^{2}-E_{nl}^{2}}\,R)} \,,
\end{equation}  
where $K_{l}(x)$ is the modified Bessel function of second kind. 

For $l=0$, quantization condition in Eq.~\eqref{eq:energylevels} gives a simple relation: $-\cot(E_{n}R)=\sqrt{m_{0}^{2}-E_{n}^{2}}/E_{n}$. Notice that this only has a solution when $m_{0}R>\pi/2$, and thus there is a threshold $R$ for the existence of at least one bound state $R>R_{\rm{th}}= \frac{\pi}{2}\,m_{0}^{-1}$. For a general $l$, the threshold for $R$ is $R>R^{l}_{\rm{th}}= J_{l-1/2, 1} m_{0}^{-1}$ with $J_{l-1/2, 1}$ as the first zero of Bessel function $J_{l-1/2}$. For instance, $J_{0-1/2,1}=\pi/2$, $J_{1-1/2,1}=\pi$, and $J_{2-1/2,1}=4.49$. For a general $R$, there exists a largest value of $l$ or $l_{\rm{max}}$, above which there are no bound states. The $l_{\rm{max}}$ is given by
\begin{equation}
    J_{l_{\rm max}- \frac{1}{2}, 1 } \le R\,m_0 < J_{l_{\rm max} + \frac{1}{2}, 1 } \, .
\end{equation}

\begin{figure}[ht!b]
\centering
    \includegraphics[width=0.6 \textwidth]{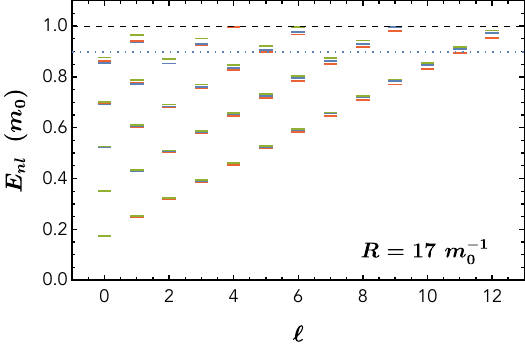}
    \caption{Example of energy levels for a fixed $R = 17\,m_0^{-1}$ with $m_0 = g\,v_n = 327$~MeV to accommodate an $A = 1000$ quark nugget (occupying the energy levels below the blue dotted line). The red-colored energy levels are from the numerical full results in \eqref{eq:energylevels}; the blue-colored ones are from using the approximate formula in \eqref{eq:En-approx}; and the green-colored ones are from using an even simpler approximate formula in \eqref{eq:En-approx-simpler}.
    }
    \label{fig:energy-level-example}
\end{figure}

To obtain analytic formulas for the energy levels, we note that the outside wave function has the following approximate formulas
\beqa
\frac{d[a^{\rm out}_{nl}(r)/r]/dr}{a^{\rm out}_{nl}(r)/r}\Big|_{r = R}  \approx - \, k \, - \frac{1}{R}  ~,
\eeqa
in the limit of $k R \gg 1$ with $k \equiv \sqrt{m_0^2 - E_{nl}^2}$. For the inside wave function and without approximation, one has 
\beqa
\frac{d[a^{\rm in}_{nl}(r)/r]/dr}{a^{\rm in}_{nl}(r)/r}\Big|_{r = R} = \frac{l}{R} \,-\, \frac{E_{nl} \, j_{l + 1}(E_{nl}\, R) }{ j_{l }(E_{nl}\, R) }  ~.
\eeqa
Equating the above two equations, the energy quantization condition becomes 
\beqa
\frac{(k \, R + l + 1) } {E_{nl} \, R }  =   \frac{J_{l + 3/2} (E_{nl} \, R) } { J_{l + 1/2}(E_{nl} \, R)  } ~.
\label{eq:En}
\eeqa
To proceed further, we need to use the Mittag-Leffler expansion~\cite{Watson:1944:TBF}
\beqa
\frac{J_{\nu + 1}(x)}{J_\nu(x)} = 2 \,x \sum^\infty_{n = 1} \frac{1}{J^2_{\nu, n}  - x^2 } ~, \qquad x\neq \pm J_{\nu, n}\,, \qquad n = 1, 2, \cdots ~,
\eeqa
with $J_{\nu, n}$ as the $n$'th zero of the Bessel function $J_\nu(x)$. Making the single-pole dominance assumption to solve Eq.~\eqref{eq:En}, the $E_{nl}$ has the approximate formula
\begin{scriptsize}
\beqa
\label{eq:En-approx}
E_{n l} \approx \frac{\sqrt{ J^2_{l + 1/2, n} + 2 m_0 R (m_0 R + l + 3) - \sqrt{ (J^2_{l + 1/2, n} + 2 m_0 R (m_0 R + l + 3) )^2 - 8 \, J^2_{l + 1/2, n}\,m_0 R (m_0 R + l + 1) }  }}{\sqrt{2}\, m_0 R }\, m_0 ~.
\eeqa
\end{scriptsize}
In the limit $m_0^{2} \gg \frac{E_{nl}^2}{2}$ or the deeper-bounded energy levels, one has a very simple formula for $E_{n l}$ as 
\beqa
\label{eq:En-approx-simpler}
E_{n l} = J_{l + 1/2, n}\, \sqrt{\frac{ l + m_0 R + 1}{ l + m_0 R + 3 } }\, \frac{1}{R} ~. 
\eeqa
In Fig.~\ref{fig:energy-level-example}, we show the energy levels for one example radius with $R = 17 \, m_{0}^{-1}$ based on the full quantization condition in \eqref{eq:energylevels} and approximate conditions in \eqref{eq:En-approx} and \eqref{eq:En-approx-simpler}. The particular choice of $R$ approximates the quark nugget radius with $A = 1000$ after minimizing $\frac{E(R)}{A}$ as a function of $R$. Note that to count the baryon number, one has $2 \cdot 2 \cdot (2l + 1)$ for each energy level, with one factor of $2$ for two spins and the other factor of $2$ for $u$ and $d$ flavors. The color factor of 3 cancels the baryon number $\frac{1}{3}$ for each quark.

\begin{figure}[th!b]
    \centering
    \includegraphics[width=0.48\textwidth]{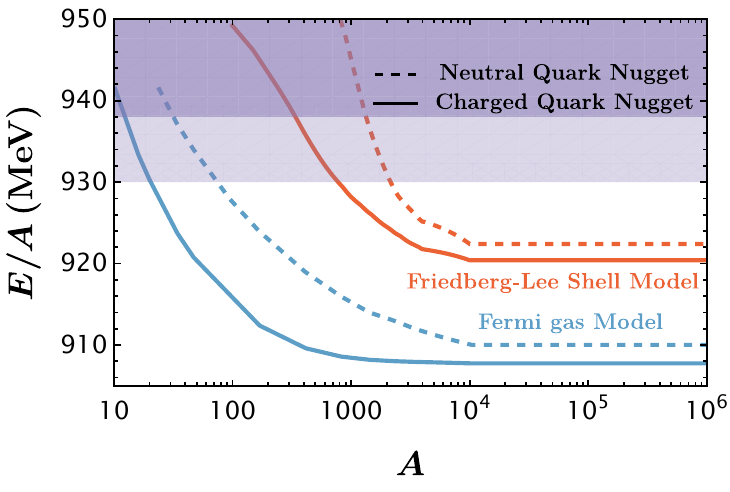}
    \hspace{3mm}
    \includegraphics[width=0.48\textwidth]{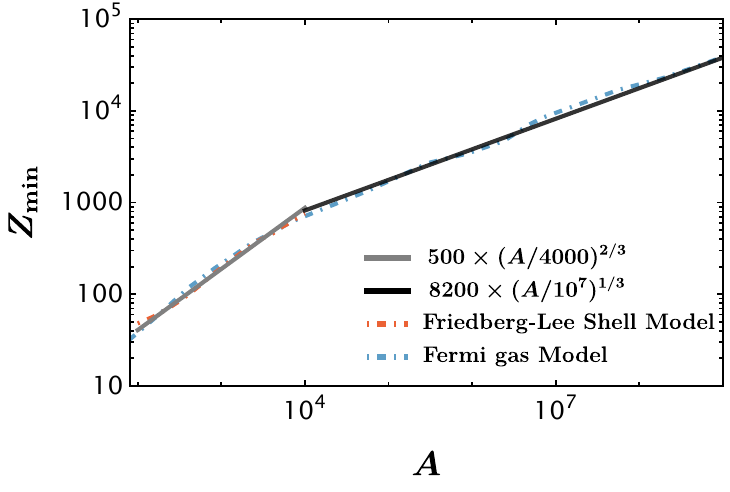}
    \caption{{\it Left panel:} The energy per baryon $E/A$ as a function of $A$ based on the degenerate Fermi gas model (blue curve) and the Friedberg-Lee Shell model (red curve). Both electrically charged (solid line) with $Z_{\rm min}$ and neutral (dashed line) quark nuggets are shown here. The two violet-colored regions denote the meta-stable regions for $E/A$ above the nucleon mass (darker) and the ion element value (lighter). {\it Right panel:} $Z_{\rm{min}}$ as a function of $A$ for the two models. A broken-power-law fit to the data is shown in solid lines.
    }
    \label{fig:EbyA_compare}
\end{figure}

Based on the calculated energy levels, we then calculate the energy $E(A, Z)$ for a given two-flavor quark nugget $\ce{^A_Z Q}$. Denoting $N_{u}, N_{d}$ as the total numbers of up and down quarks respectively, we have $N_{u}+N_{d}=3\,A$ and $2N_{u}-N_{d}=3\,Z$ or $\{N_u, N_d\} = \{A+Z, 2 A - Z\}$. For a given $R$, we fill $N_u$ and $N_d$ numbers of energy levels including the spin and $2l + 1$ factors to obtain the total Fermi energy as a function of $R$. After adding the vacuum energy and the Coulomb energy [simply taken to be $3\alpha Z^2/(5 R)$], we minimize the total energy in $R$ to obtain the minimum energy at a $R_{\rm min}$. After calculating $E(A, Z)$, we can identify a $Z_{\rm min}$ for the minimum of $E(A, Z)$ with a fixed $A$. The value of $E(A, Z_{\rm min})/A$ as a function of $A$ is shown in the solid red line in the left panel of Fig.~\ref{fig:EbyA_compare}. As a comparison, we also show the $E/A$ of electrically-neutral quark nuggets in the dotted red line, which has a higher value of $E/A$ by a few MeV. 

As a consistence check of our step-function assumption for the scalar field profile, we can estimate a naive ``Fermi momentum" based on the baryon number $A$ and $R_{\rm min}$ or $p_F^{\rm naive} = (3\pi^2\,n_A)^{1/3} = (9\pi/4)^{1/3}\,A^{1/3}\,R^{-1}_{\rm min}$. For $A=1000$ and $R_{\rm min} = 17\,m_0^{-1}$, one has $p_F^{\rm naive} = 368.7$~MeV, which is large enough to reduce the $\sigma_n$ field close to zero (see the left panel of Fig.~\ref{fig:bulk-energy}). 

Comparing the curves based on the two models: ``Fermi gas model" and ``Friedberg-Lee Shell Model", one can see that the latter has a larger value of $E/A$. As we discussed before, the Friedberg-Lee Shell Model provides a more precise modeling for a quark nugget than the Fermi gas model. One simple reason is that the quark energy levels shown in Fig.~\ref{fig:energy-level-example} have a discrete feature and do not match the simple Fermi gas model with a simple power-law behavior for the density of states in energy.

For both models, the $E/A$ approaches a constant value at large $A$ or the bulk limit. For the benchmark model, the Friedberg-Lee Shell Model has $920$~MeV in the large $A$ limit, while the Fermi gas model has $908$~MeV. Comparing to the value of the iron element ($E/A$ around 930~MeV), the (quantum) stability value for $A_{\rm min}$ is 800 for the charged quark nuggets and 2100 for the neutral nuggets in the Friedberg-Lee Shell Model. In contrast, the corresponding values are 20 and 81 for the Fermi gas model. However, we note that the value of $A_{\rm min}$ should not be trusted based on the Fermi gas model or the calculation using the step-function approximation for the Friedberg-Lee Shell Model. We will calculate this value more precisely in the next subsection.

In the right panel of Fig.~\ref{fig:EbyA_compare}, we show $Z_{\rm{min}}$ of the charged quark nuggets as a function of $A$ for both the Friedberg-Lee Shell model (red dot-dashed curve) and the Fermi gas model (blue dot-dashed curve). One can see that both models have approximately the same dependence. We also provide a broken power-law fit for the behavior. In the large $A$ limit above $10^4$, the dependence matches well to the simple formula derived in Eq.~\eqref{eq:bulk-Z-min}, which serves as a consistent check.  

\subsubsection{Solving Fermion and scalar coupled EOM's}
\label{sec:coupled-EOM}

In this section, instead of making the step-function assumption for the $\sigma_n$ field, we solve the coupled EOMs for the fermion and scalar fields. Although it is more challenging to solve two coupled second-order (or one second-order plus two first-order) differential equations, the quark nuggets with a smaller baryon number can be modeled more precisely. Earlier studies solving the coupled EOMs based on the Friedberg-Lee Shell Model can be found in Ref.~\cite{Koeppel:1985tt}.

The scalar equation of motion can be found in \eqref{eq:sigma-EOM-surface} after replacing the last term by $g\,N_l \langle\bar{\psi} \psi\rangle$ with $N_l$ as the number of quarks in the specific partial wave number $l$~\cite{Friedberg:1977xf}. Using the partial-wave expansion in \eqref{eq:fermion-decomposition} and the following re-scaling 
\beqa
    r &\rightarrow & \bar{r}/(g v_{n})\,, \quad \quad \sigma_{n}\rightarrow \bar{\sigma}_{n}\,v_{n}, \quad \quad E_{0}\rightarrow \overline{E}_{0}\,g\,v_{n}\,, \nonumber \\ \, V&\rightarrow &\overline{V}\,g^{2}\,v_{n}^{4}\,, \quad \quad (a_{0},b_{0})\rightarrow [v_{n}/(g N_l)]^{1/2}(\bar{a}_{0},\bar{b}_{0})\,,
\eeqa
the $s$-wave equations of motion for the Fermion and scalar fields are
\beqa
\label{eq:coupled-shell-model-EOM}
&& \frac{d\bar{a}_{0}}{d\bar{r}} \,=\, \frac{\bar{a}_{0}}{\bar{r}}  + [\bar{\sigma}_{n}(\bar{r}) + \overline{E}_{0}]\,\bar{b}_{0} ~, \qquad
\frac{d\bar{b}_{0}}{d\bar{r}} \,=\,   [\bar{\sigma}_{n}(\bar{r})- \overline{E}_{0}]\,\bar{a}_{0} - \frac{ \bar{b}_{0}}{\bar{r}} ~,\nonumber\\
&& \frac{d^{2}\bar{\sigma}_{n}}{d\bar{r}^{2}} + \frac{2}{\bar{r}}\frac{d\bar{\sigma}_{n}}{d\bar{r}} \,=\, \frac{d\overline{V}}{d\bar{\sigma}_{n}} + \frac{1}{\bar{r}^{2}}(\bar{a}_{0}^{2} - \bar{b}_{0}^{2}) ~.
\eeqa
The Fermion fields $\bar{a}_0$ and $\bar{b}_0$ satisfy the normalization condition: $4\pi \int_{0}^{\infty} d\bar{r} \, (\bar{a}_{0}^{2} + \bar{b}_{0}^{2}) = \bar{\eta} \equiv N_{l=0} g^{2}$. The boundary conditions are
\begin{equation}
    \bar{a}_{0}(\bar{r}=\infty)=0, \quad \bar{b}_{0}(\bar{r}=0)=0, \quad \bar{\sigma}_{0}(\bar{r}=\infty)=1, \quad \frac{d\bar{\sigma}_{0}}{d\bar{r}}(\bar{r}=0) = 0\,.
\end{equation}
Using the shooting method to solve the boundary condition problems, we obtain both the eigenvalue $\overline{E}_0$ and the wave-functions $\bar{a}(\bar{r})$, $\bar{b}(\bar{r})$ and $\bar{\sigma}_n(\bar{r})$. We can then calculate the total energy $E$ of the quark nugget by summing the Fermion and scalar energies 
\beqa
    \overline{E} \equiv \frac{g\, E}{v_{n}} = \bar{\eta}\, \overline{E}_{0} + 4 \pi \int_{0}^{\infty} d\bar{r} \, \bar{r}^{2} \left[ \frac{1}{2} \left( \frac{d\bar{\sigma}_{n}}{d\bar{r}} \right)^{2} + \overline{V}(\bar{\sigma}_{n}) \right] \, .
\eeqa

\begin{figure}[th!b]
    \centering
    \begin{subfigure}{0.44\textwidth}
        \centering
        \includegraphics[width=\linewidth]{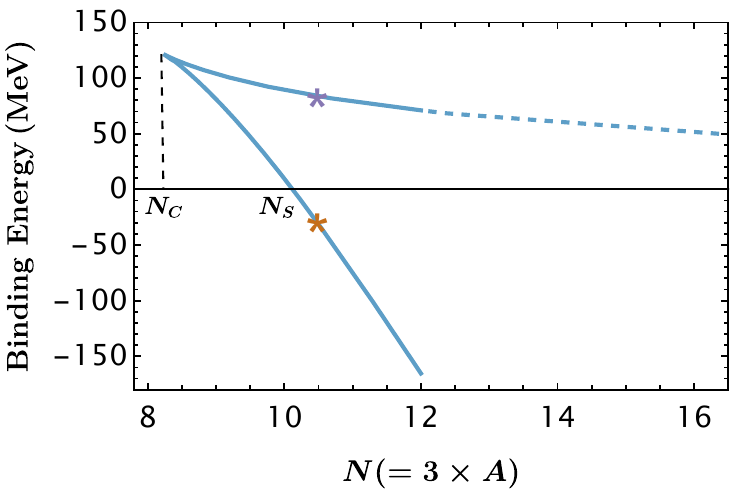} 
    \end{subfigure}%
    \hspace{3mm}
    \begin{subfigure}{0.44\textwidth}
        \centering
        \includegraphics[width=\linewidth]{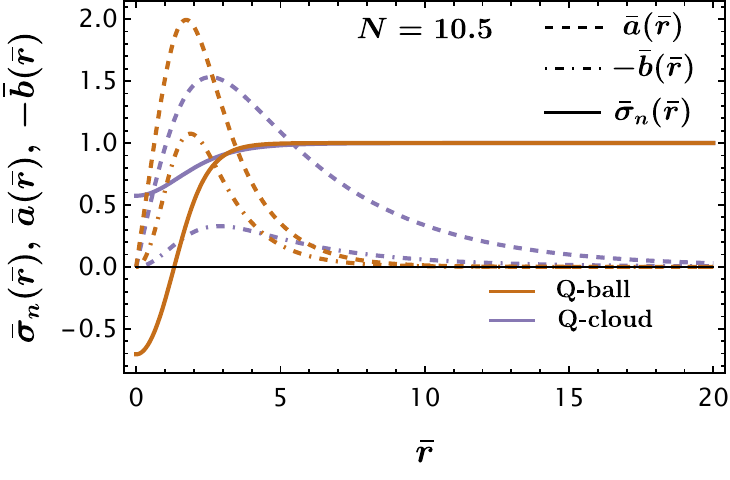} 
    \end{subfigure}
    \caption{{\it Left panel}: The binding energy $E - N\,g\,v_n$ of quark nuggets as a function of the number of quarks $N$ ($N \leq 12$ for fermions in $s$-wave). The baryon number $A = N /3$. $N_S$ (around 10) is the minimum $N$ to have stable quark nuggets against decaying into free nucleons at the quantum level. {\it Right panel}: Example profiles of $\bar{a}(\bar{r})$ (dashed), $-\bar{b}(\bar{r})$ (dot-dashed), and $\bar{\sigma}_{n}(\bar{r})$ (solid) for both Q-ball (brown) and Q-cloud (purple) solutions with the same $N=10.5$ (labeled by the brown and purple stars in the left panel).
    }
    \label{fig:Qnugget_Nc}
\end{figure}

In the left panel of Fig.~\ref{fig:Qnugget_Nc}, we show the binding energy of the quark nuggets, given by $E- N\,g\,v_{n}$, as function of $N$. For $s$-wave quarks, one has $N = N_{l = 0}$. Because of the Pauli exclusion principle, the maximum number of quarks to occupy the $s$-wave orbits is $2\times 2 \times 3 = 12$. Therefore, our calculations based on \eqref{eq:coupled-shell-model-EOM} can only be applied to $N_{l=0} = 12$ or the baryon number $A = N_{l = 0} /3 = 4$. To go beyond $A=4$, one also needs to include higher-partial-wave states like $p$-, $d$-, or $f$-wave states.  

As can be seen in this plot, there are two branches of solutions for the system, which is a generic feature for non-topological solutions in $3+1$ dimensions~\cite{Friedberg:1976me, Koeppel:1985tt, Lee:1991ax, Levkov:2017paj, Bai:2022kxq}. The lower branch is the so-called ``Q-ball" branch, while the upper branch is the ``Q-cloud" branch and is classically unstable. The lower and upper branches intersect at $N = N_C$ (around 8), which is the smallest fermion number to have a soliton solution. For $N > N_S$ (around 10), the binding energy is negative, such that the quark nuggets are stable at the quantum level (their energy is smaller than the summed energy of free massive quarks with the same $A$). For the range $N_C < N < N_S$, the quark nugget (a type of Fermionic non-topological soliton) is classically stable but metastable at the quantum level. Through quantum tunneling, similar to $\alpha$ decay in some nuclei, the quark nuggets can evaporate into $N$ free quark states or $A = N/3$ free nucleons, to be more precise.  

In the right panel of Fig.~\ref{fig:Qnugget_Nc}, we show the profiles of both Fermion and scalar fields with $\bar{\sigma}_{n}(\bar{r})$ (solid), $\bar{a}(\bar{r})$ (dashed), and $\bar{b}(\bar{r})$ (dot-dashed). We present two examples: one for the Q-ball solution (in brown) with $\bar{E}_0 = 0.74$ and one for the Q-cloud solution (in blue) with $\bar{E}_0 = 0.97$. Both examples have $N=10.5$ to demonstrate the profile differences between these two types of solutions. For instance, the $\bar{\sigma}_n(\bar{r})$ profile for the Q-cloud solution is flatter than that for the Q-ball solution. We also note that the $\bar{\sigma}_n(\bar{r})$ profile for the Q-ball solution has a negative value at the origin. This can be understood from the potential $V[\sigma_n, v_s]$ in the left panel of Fig.~\ref{fig:bulk-energy} for $p_F = 0$. There is another local minimum of the potential at $\sigma_n \approx -80$~MeV or $\bar{\sigma}_n \equiv \sigma_n/v_n \approx -0.87$. The Fermion energies provide an additional $\sigma_n$ dependence, shifting the value of $\bar{\sigma}_n$ slightly above $-0.87$.    

\section{Radioactivity of quark nuggets}
\label{sec:radioactivity}
The ordinary nuclei, $\ce{^A_Z N}$, can exhibit various types of radioactivity, including $\gamma$ decay, $\beta$ decay, $\alpha$ decay, spontaneous fission, and neutron and proton emission. Many properties, such as lifetimes, transition energies, and branching ratios, are known from theoretical calculations or experimental measurements~\cite{Kondev:2021lzi}. In contrast, the radioactivity properties of quark nuggets, $\ce{^A_Z Q}$, are not only less known but also not well studied. In this section, we examine some of their radioactivity properties based on the quark nugget properties calculated in the previous section. As emphasized before, both the degenerate Fermi gas model and the Friedberg-Lee Shell model are simplified phenomenological models and should not be taken at face value. However, some general features of quark nugget radioactivity can be learned and can be adopted to search for quark nuggets. 

We first notice that the $E/A$ for quark nuggets could be $\mathcal{O}(10~\mbox{MeV})$ smaller than the free nucleon mass (even compared to the iron nucleus). For the degenerate Fermi gas model, the difference $\Delta(E/A) \equiv m_p - E/A|_{\rm quark\, nugget} \approx 30$~MeV (see Fig.~\ref{fig:EbyA_compare}). This means that the free proton or neutron emission of quark nuggets requires a much higher excited state compared to ordinary nuclei, which have a threshold excitation energy of 12~MeV or below~\cite{nucleardata}. Since the decay width for free nucleon emission is $10-100$~keV, the electromagnetic decay branching ratios could be larger for some excited quark nugget states inaccessible to the free nucleon emission channel.

\subsection{Gamma decay}
\label{sec:gamma}

For an excited quark nugget, it can de-excite to various lower states by emitting photons
\beqa
\ce{^A_ZQ}^{***} \rightarrow \ce{^A_ZQ}^{**} + \gamma_1 \rightarrow \ce{^A_ZQ}^{*} + \gamma_1 + \gamma_2 \rightarrow  \ce{^A_ZQ} + \gamma_1 + \gamma_2 + \gamma_3~,
\eeqa
with the gamma-ray energy depending on the energy levels of transitioning states. Here, we use an example of cascading decays involving three photons to demonstrate the abundance of gamma rays from excited quark nuggets. For ordinary nuclei or atomic states, one could use the photon energy or spectral line to label or identify the nuclei. Similarly, quark nuggets $\ce{^A_ZQ}$ with different $A$ and $Z$ could have their own spectral-line or gamma-ray line labeling. Unlike ordinary nuclei, which have the heaviest atomic number and mass up to $\ce{^{294}_{118}Og}$~\cite{Oganessian:2006va}, quark nuggets could have a much larger atomic mass, leading to more excited energy levels and denser spectral lines.

\begin{figure}[htb!]
    \centering
    \begin{subfigure}{0.8\textwidth}
        \centering
        \includegraphics[width=\linewidth]{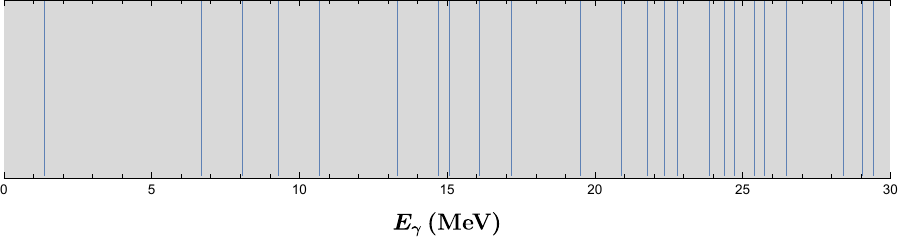} 
    \end{subfigure}%
    \vspace{3mm}
    \begin{subfigure}{0.8\textwidth}
        \centering
        \includegraphics[width=\linewidth]{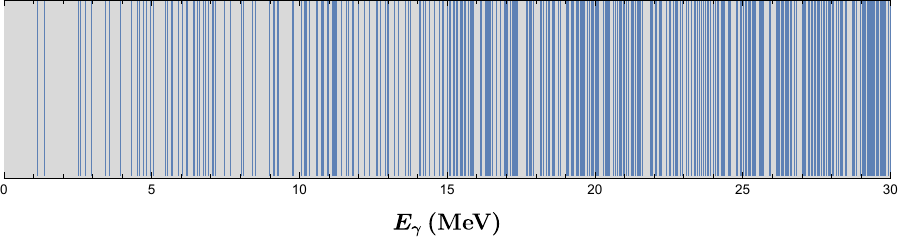} 
    \end{subfigure}
     \caption{The transition gamma-ray lines from excited bound states to ground states. The upper one is for $A=1000$ and $R=17\,m_0^{-1}$, while the lower one is for $A=10^4$ and $R=37\,m_0^{-1}$.}
    \label{fig:gamma-spectrum}
\end{figure}
 
It is not feasible to calculate and present all spectral lines for different quark nuggets. Here, we choose two examples of quark nuggets with $A=1000$ ($R=17\,m_0^{-1}$) and $A = 10^4$ ($R=37\,m_0^{-1}$) to demonstrate the general features of gamma-ray spectral lines in Fig.~\ref{fig:gamma-spectrum}. In this figure, we only show the transition energies or the corresponding gamma-ray energies. In principle, different spectral lines could have different intensities, which depend on the transition energies as well as the transition types. Comparing the two spectra in this figure, it is clear that a larger $A$ quark nugget has denser gamma-ray spectral lines. This feature could serve as a characteristic one to distinguish quark nuggets from ordinary nuclei.

\subsection{Beta decay and electron capture}
\label{sec:beta}

\begin{figure}[thb!]
    \centering
    \begin{subfigure}{0.44\textwidth}
        \centering
        \includegraphics[width=\linewidth]{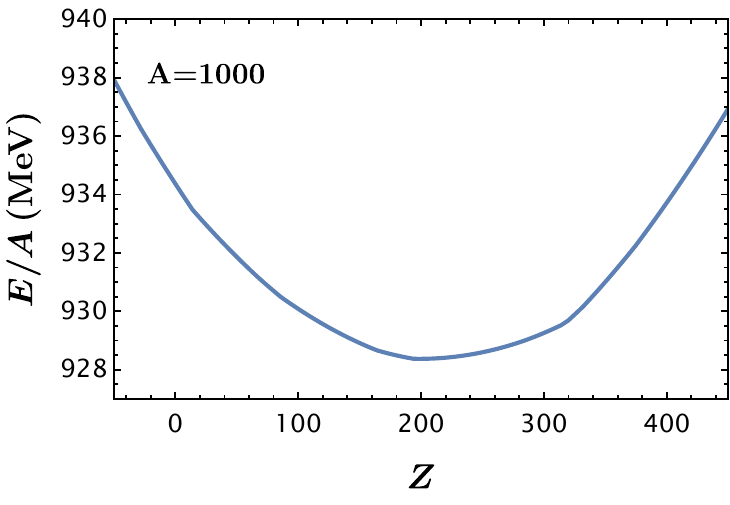} 
    \end{subfigure}%
    \hspace{6mm}
    \begin{subfigure}{0.44\textwidth}
        \centering
        \includegraphics[width=\linewidth]{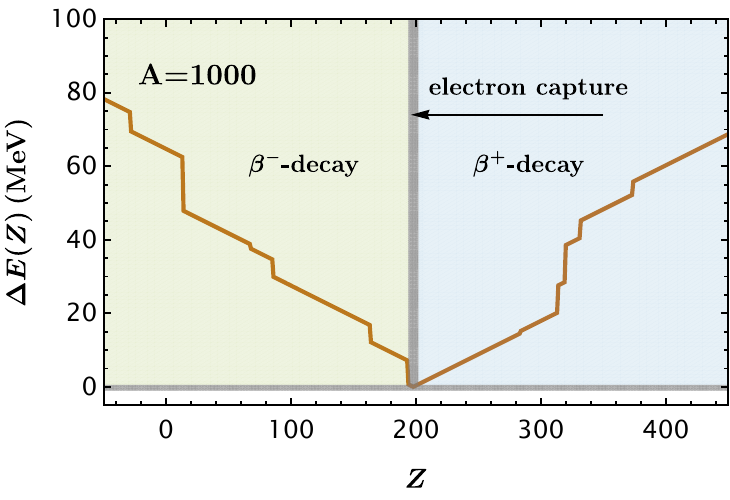} 
    \end{subfigure}
    \caption{{\it Left panel}: $E/A$ (in MeV) for different $Z$ values for $A=1000$. The $E/A$ has a minimum at $Z_{\text{min}}=198$. {\it Right panel}: $\Delta E(Z)$ as a function of $Z$. When $\Delta E(Z) > m_{e}$, the nugget is unstable against $\beta$ decay. The left green region with $Z<195$ has $\beta^{-}$ decay, while the right blue region with $Z>201$ has $\beta^{+}$ decay. The gray shaded region with $195 \leq Z \leq 201$ indicates the region of stable quark nuggets. For $Z > 195$, the quark nuggets could also undergo the electron capture process. 
    }
    \label{fig:betadecay}
\end{figure}

For a given $A$, there exists a $Z_{\rm min}$ with the lowest mass. Using $A = 1000$ as an example and in the left panel of Fig.~\ref{fig:betadecay}, one has $Z_{\rm min} = 198$ based on the Friedberg-Lee Shell model. In the right panel of Fig.~\ref{fig:betadecay}, we show the mass difference between two nearby $Z$'s. When $Z > Z_{\rm min}$ and $\Delta E(Z) \equiv E(Z)-E(Z-1) > m_e$ or $Z > Z_{\beta^+}$, this quark nugget can undergo a $\beta^+$ decay via 
\beqa
\ce{^A_ZQ} \rightarrow \ce{^{A}_{Z-1}Q} + e^+ + \nu_e ~.
\eeqa
In the other direction with $Z < Z_{\rm min}$ and $\Delta E(Z) \equiv E(Z)-E(Z+1) > m_e$ or $Z < Z_{\beta^-}$, it can undergo a $\beta^-$ decay via 
\beqa
\ce{^A_ZQ} \rightarrow \ce{^{A}_{Z+1}Q} + e^- + \bar{\nu}_e ~.
\eeqa
In the green region of Fig.~\ref{fig:betadecay} for $Z < Z_{\beta^-} = 195$, the quark nuggets have $\beta^{-}$ decay, while for $Z > Z_{\beta^+} = 201$ the quark nuggets have $\beta^{+}$ decay. The quark nuggets with $Z_{\beta^-} \leq Z \leq Z_{\beta^+}$ are therefore stable. 

For a charged quark nugget, additional electrons are likely to move around it to form an atomic state. For $Z > Z_{\beta^-}$, one could also have the electron-capture process
\beqa
\ce{^A_ZQ} + e^- \rightarrow \ce{^{A}_{Z-1}Q}  + \nu_e ~,
\eeqa
which can further reduce the quark nugget electric charge and make the electric charge closer to $Z_{\beta^-}$.

For both beta decays and the electron-capture process, electron (anti-)neutrinos are universal products with their energies of $\mathcal{O}(1\,\mbox{MeV})$ to $\mathcal{O}(10\,\mbox{MeV})$, which could be used to search for quark nuggets. Specifically, for the $\beta^+$ decay, the decay-generated positrons could collide with electrons in the medium and pair-annihilate into two gamma rays. If the positrons lose a significant fraction of their energies before the pair-annihilation process, the two gamma ray energies could just match the electron mass, or 511 keV.   

\subsection{Spontaneous fission}
\label{sec:fission}

As a subclass of non-topological solitons, quark nuggets usually have two critical baryon numbers associated with their stability. The first one is $A_C = N_C/3$ (see the left panel of Fig.~\ref{fig:Qnugget_Nc} for an example), below which there is no isolated soliton solution~\cite{Lee:1991ax}. The second one is $A_S = N_S /3$, or the critical baryon number to be stable at the quantum level. When $A > A_S$, the quark nuggets have a smaller value of energy per baryon than a free nucleon, hence they are stable against free nucleon emission.

For the baryon number between $A_C$ and $A_S$ or $A_C \leq A < A_S$, the quark nuggets are classically stable, but not stable at quantum level. If energetically allowed, it could decay into a light quark nugget plus an ordinary nucleus: $\ce{^A_ZQ} \rightarrow \ce{^{A-A_1}_{Z-Z_1}Q} + \ce{^{A_1}_{Z_1}N}$. Alternatively, the quark nuggets could also ``evaporate" or ``explode" into $A$ free nucleons
\beqa
\ce{^A_ZQ} \xrightarrow[]{\text{explosion}} \underbrace{p + p + \cdots + p}_Z + \underbrace{n + n + \cdots + n}_{A-Z} ~,
\eeqa
which could also generate additional gamma rays from charged-particle bremsstrahlung or the binding of nucleons to form ordinary nuclei. 

Because of the quantum-tunneling nature of the explosion process, the decay width could be exponentially suppressed. As a result, the quark nugget with $A_C \leq A < A_S$ could have a long lifetime and a delayed decay process. Based on the semiclassical method and a Q-ball model with a self-interacting potential of a single complex field, Ref.~\cite{Levkov:2017paj} provides a general formula 
\beqa
\Gamma_Q \approx m_0 \, A_C^{1/2}\,\exp\Big[- 3\,(A - A_C) \big(c_1 + c_2 \ln{(1 - A/A_S)}\big) \Big]~,
\eeqa
with $c_1, c_2 < 0$ ($c_1 = -0.28$ and $c_2 = -2.6$ for the Q-ball model in Ref.~\cite{Levkov:2017paj}; see also Ref.~\cite{Son:2021kkx} for calculations based on a non-relativistic system). Although we anticipate different model-dependent $\mathcal{O}(1)$ numbers for $c_1$ and $c_2$ for the particular Friedberg-Lee Shell model, the decay width of quark nuggets should contain a similar exponential factor, or $\Gamma_Q \propto m_0\,e^{-c\,(A - A_C)}$ with $c$ as an order-unit number.

From the left panel of Fig.~\ref{fig:Qnugget_Nc}, we have $A_C = 2$ and $A_S = 4$, such that the quark nugget with $A = 3$ or $N = 9$ undergoes an explosion process to decay into three ordinary nucleons. We note that the calculation in Fig.~\ref{fig:Qnugget_Nc} is based on $s$-wave quark states. One could extend the calculation to higher partial waves, which may have multiple instability intervals in $A$ leading to the explosion process. Furthermore, the linear sigma model studied here is intended to provide guidance on some generic features of quark nuggets. A non-perturbative tool may be necessary to precisely calculate values like $A_C$ and $A_S$. Because of the exponential factor, the quark nugget's lifetime based on the spontaneous fission process could be much longer than those for gamma and beta decays. Some quark nuggets could have decay lifetimes on a longer time scale, such as minutes, hours, or even days.

\section{Discussion and conclusions}
\label{sec:discussion}

In this paper, we have employed the linear sigma model with quarks to study the properties of quark nuggets. We have considered two distinct cases: the degenerate Fermi gas model, which describes the general behavior of quark nuggets with large baryon numbers, and the Friedberg-Lee shell model, which offers a better approximation for quark nuggets with lower baryon numbers. In both models, we account for quark kinetic energy, vacuum and surface energy contributions from scalar fields, and electromagnetic interaction energy. However, we do not include the effects of QCD gluon exchange for the quark gas. Historically, the strong interaction effects in the bag model could be incorporated through $\mathcal{O}(\alpha_{s})$ corrections to the thermodynamic potentials, which could alter the stability criteria for quark nuggets. This correction could be absorbed into a modification for the bag parameter, with larger $\alpha_{s}$ values favoring nugget stability at smaller bag parameters and instability at larger ones~\cite{Farhi:1984qu}. Although we use a different underlying model from Ref.~\cite{Farhi:1984qu}, we anticipate that including gluon exchange would similarly affect the quark nugget stability and the effect could be absorbed with modifications of model parameters in the linear sigma model.

In addition to studying neutral quark nuggets, we have also investigated charged quark nuggets. In the large baryon number limit, stable quark nuggets exhibit a charge proportional to $Z \propto A^{1/3}$, which implies that their charge-to-mass ratio could be significantly smaller than that of ordinary nuclei. Similar to ordinary nuclei, charged quark nuggets are expected to be surrounded by a cloud of electrons, forming a neutral ``quark nugget atom." Depending on the quark nugget's charge and radius, the electrons may either reside within the nugget’s bulk, described by an electron density profile, or outside the nugget in atomic orbitals.

By comparing the (non-relativistic) electron Bohr radius with the quark nugget radius, we find that for $A \lesssim 10^4$ or $Z \lesssim 10^3$, the electrons are located outside the quark nugget core. In this case, one can use the shell model to calculate the atomic energy levels of the quark nugget atom. However, relativistic effects become significant around $Z \approx 170$, complicating the energy calculations~\cite{ElectronLargeZ,ElectronReview}. Since ordinary nuclei have atomic numbers up to 118, observing spectral lines from an atom with $118 < Z < 10^3$ could serve as a ``smoking gun" for identifying the existence of quark nuggets.

Even for smaller values of $Z$, both the ordinary nucleus $\ce{^A_Z N}$ and the quark nugget $\ce{^{A'}_Z Q}$—with different values of $A$ and $A'$—could exist and be long-lived. The difference in their charge distributions would result in distinct energy levels and atomic spectral lines, which could be experimentally distinguished. For very large values of $Z$, a crude Thomas-Fermi model can be applied to calculate the general electron distribution and the resulting continuous thermal radiation~\cite{Flambaum:2021xub,Flambaum:2021awu}.

A potential observational site to detect quark nuggets could be in the post-merger remnants of neutron or (strange) quark star collisions, which are predicted to produce quark nuggets. To observe the radioactive signatures of quark nuggets in situ, it is also necessary to know the baryon-number distribution of the produced quark nuggets. This distribution depends on the fragmentation models, as discussed in Refs.~\cite{Paulucci:2014vna,Bucciantini:2019ivq}, although a more thorough study is required. By convolving the baryon-number distribution with the spectral lines from radioactivity or atomic transitions, one could develop a strategy to search for quark nuggets from binary neutron star mergers.

In addition to their potential formation during neutron star or (strange) quark star mergers, quark nuggets could also have formed in the early universe from phase transitions or domain wall collapses related to the QCD scale. These nuggets, which could constitute all or part of dark matter, may exist in significant abundance within our galaxy. Cosmic rays interacting with the nuggets could excite them, resulting in higher atomic numbers or increased charge states. The subsequent de-excitation of these nuggets, through radioactive decays or atomic transitions, could produce unique spectral lines. If quark nuggets exist in an excited, charged state, their de-excitation might involve $\beta^{+}$ decay, where emitted positrons would lose energy and annihilate with background electrons, generating the 511 keV gamma-ray line. This line has been observed from the Galactic Center and could serve as an additional probe for quark nuggets~\cite{Prantzos:2010wi}.

Quark nuggets present in our galaxy are expected to propagate and potentially reach Earth~\cite{Madsen:2004vw}. Since quark nuggets can be modeled as cosmic rays with a higher atomic number and lower $Z/A$ ratio compared to ordinary nuclei, they may produce distinct signals in cosmic ray detectors and have already been searched for at the AMS-02~\cite{AMS:2021nhj}. A similar study could also be conducted at the DAMPE experiment~\cite{DAMPE:2017cev}. Additionally, a weaker limit on their flux has already been derived using samples from lunar soil~\cite{Han:2009sj}. For direct detection experiments, if quark nuggets penetrate the overburden material, they provide an interesting multi-hit signature that could be searched for in large volume detectors~\cite{Bramante:2018qbc,Bai:2022nsv}. The new radioactivity signatures presented in this paper could refine these search limits and provide new search strategies, such as using meteor detectors~\cite{Dhakal:2022rwn} or searching for gamma-ray bursts in the time domain~\cite{Kaplan:2024dsn}.  

In conclusion, we have utilized the linear sigma model with constituent quarks, incorporating both the degenerate Fermi gas and Friedberg-Lee shell models, to characterize quark nugget properties at high and low baryon number concentrations. Our findings suggest that stable quark nuggets consist solely of up and down quarks for both neutral and charged cases, with charged nuggets exhibiting slightly greater energetic stability compared to their neutral counterparts. The optimal charge that minimizes the energy scales as $A^{1/3}$ for larger quark nuggets. The mass per baryon of quark nuggets could be $10$-$20$~MeV smaller than that of iron. Additionally, we have identified potential radioactivity signatures for quark nuggets, including emissions through gamma-ray emission, beta decay, and spontaneous fission. Some gamma-ray spectral lines could provide a ``smoking gun" to search for quark nuggets. We also point out that a quark nugget, with a baryon number above the classical stable one but below the quantum stable value, could have a delayed explosive spontaneous fission. These radioactivity properties could guide the search for quark nuggets, including from binary neutron star mergers.

\subsubsection*{Acknowledgments}
We thank Dean Chen and Raymond E. Frey for useful discussion. The work of YB is supported by the U.S. Department of Energy under the contract DE-SC-0017647. MK acknowledges support from the National Science Foundation with the Grant No. PHY-2020275.

\appendix
\section{The linear sigma model}
\label{sec:LMSq}

In this appendix, we provide the details of the linear sigma model which is used to derive the properties of quark nuggets in Section~\ref{sec:basis}. The linear sigma model is an effective phenomenological model that is used to derive the mass spectra and some decay rates of scalar and pseudo-scalar nonets of sub-GeV mesons. The Lagrangian, without including quark degrees of freedom, is given by 
\begin{equation}
    \mathcal{L}_{\Sigma} = \mbox{Tr}(\partial_{\mu}\Sigma^{\dagger}\partial^{\mu}\Sigma) - V(\Sigma) \,.
\end{equation}
Here, $\Sigma=T_{a}(\sigma_{a}+ i \pi_{a})$ is meson field. $T_a = \Lambda_a/2$ with $a=0, \cdots,8$ are generators of $U(3)$ and satisfy $\mbox{Tr}(T_a T_b)=\delta_{ab}/2$. $\Lambda_{a}$ for $a=1,\cdots,8$ are Gell-Mann matrices and $\Lambda_0 = \sqrt{\frac{2}{3}}\mathbbm{I}_3$. The potential $V(\Sigma)= V_{\rm{inv}}(\Sigma) + V_{\rm{b}}(\Sigma)$ has $SU(3)$ flavor preserving term $V_{\rm{inv}}$ given by
\begin{equation}
    V_{\rm{inv}}(\Sigma) = \lambda_{1}\, \big(\mbox{Tr}[\Sigma^{\dagger}\Sigma]\big)^{2} + \lambda_{2} \,\mbox{Tr}\big[(\Sigma^{\dagger}\Sigma)^{2}\big] + m^{2} \,\mbox{Tr}(\Sigma^{\dagger}\Sigma) - \kappa \big(\mbox{det}(\Sigma) + h.c.\big) \,,
\end{equation}
and explicit flavor-breaking terms are given by $V_{b}=\sum_{i=1}^{8}V_{bi}$ with
\begin{align}
    V_{b1} &= c_{1}\,\mbox{Tr}(\Sigma^{\dagger}\mathcal{M} + h.c.) \nonumber \,,\\
    V_{b2} &= c_{2}\,\epsilon_{ijk}\,\epsilon_{mnl}\,\mathcal{M}_{im}\Sigma_{jn}\Sigma_{kl} + h.c.\nonumber \,,\\
    V_{b3} &= c_{3}\, \mbox{Tr}(\Sigma^{\dagger}\Sigma \Sigma^{\dagger}\mathcal{M}) + h.c. \nonumber \,,\\
    V_{b4} &= c_{4}\, \mbox{Tr}(\Sigma^{\dagger} \Sigma) \,\mbox{Tr}(\Sigma^{\dagger}\mathcal{M}) + h.c.\nonumber \,,\\
    V_{b5} &= c_{5}\, \mbox{Tr}(\Sigma^{\dagger}\mathcal{M}\Sigma^{\dagger}\mathcal{M}) + h.c. \nonumber \,,\\
    V_{b6} &= c_{6}\, \mbox{Tr}(\Sigma \Sigma^{\dagger} \mathcal{M} \mathcal{M}^{\dagger} + \Sigma^{\dagger} \Sigma \mathcal{M}^{\dagger} \mathcal{M}) \nonumber \,,\\
    V_{b7} &= c_{7}\, (\mbox{Tr}\, \Sigma^{\dagger}\mathcal{M} + h.c.)^{2} \nonumber \,,\\
    V_{b8} &= c_{8}\, (\mbox{Tr} \,\Sigma^{\dagger}\mathcal{M} - h.c.)^{2}\,.
\end{align}
Here, $\mathcal{M}=(m_{u0},m_{d0},m_{s0})$ is the quark mass matrix, and we take  $m_{u0}=m_{d0}=3.5 \, \rm{MeV}$, $m_{s0}=96 \, \rm{MeV}$ for the bare quark masses, or the $\overline{\mbox{MS}}$ masses at the renormalization scale $\mu = 2$~GeV.  

The chiral symmetry breaking leads to vacuum expectation values for neutral meson fields at the potential minima given by $\langle \Sigma \rangle = v_{0}\,T_{0}  + v_{8}\,T_{8}  = \frac{1}{2}(v_{n},v_{n},\sqrt{2} v_{s})$. Here, we define a basis with non-strange and strange mesons as $\sigma_{n}=\frac{\sqrt{2}}{\sqrt{3}}\sigma_{0} + \frac{1}{\sqrt{3}} \sigma_{8}$, and  $\sigma_{s}=\frac{1}{\sqrt{3}}\sigma_{0} - \frac{\sqrt{2}}{\sqrt{3}} \sigma_{8}$, respectively. Using the partially conserved axial current relation~\cite{Lenaghan:2000ey}, one has  
$v_{n}=f_{\pi}=92\,\rm{MeV}$ and $v_{s}=\sqrt{2}f_{K}-f_{\pi}/\sqrt{2} = 90.5\, \rm MeV$.  

Following the procedure from Ref.~\cite{Holdom:2017gdc}, we determine the 12 model parameters ($\lambda_{1}, \lambda_{2}, \kappa, m^{2}, c_{1, \cdots, 8}$) with their values given by
\begin{align}
\label{eq:benchmark-model}
    \lambda_{1} &=-9.27, \, \lambda_{2}= 52.16,\, m^{2}=-(422.94 \, \mbox{MeV})^{2},\, \kappa=4836.29 \,\mbox{MeV}, \, c_{1}= -466816 \, \mbox{MeV}^{2},\nonumber \\ c_{2}&=221.473\,\mbox{MeV},\, 
    c_{3}=-52.46,\, c_{4}=31.94,\, c_{5}=-2.98,\, c_{6}=49.85,\, c_{7}=1.35,\, c_{8}=4.73.
\end{align}

\newpage
\setlength{\bibsep}{6pt}
\bibliographystyle{JHEP}
\bibliography{QuarkNuggetsNS}

\end{document}